\magnification=\magstep1 
\font\bigbfont=cmbx10 scaled\magstep1
\font\bigifont=cmti10 scaled\magstep1
\font\bigrfont=cmr10 scaled\magstep1
\vsize = 23.5 truecm
\hsize = 15.5 truecm
\hoffset = .2truein
\baselineskip = 14 truept
\overfullrule = 0pt
\parskip = 3 truept
\def\frac#1#2{{#1\over#2}}

\nopagenumbers
%This command suppresses the printing of page numbers.
%You should number the pages with blue pencil in upper right corner. 
%
%THE FOLLOWING THREE COMMANDS LEAVE SOME SPACE AT THE TOP OF THE LEAD PAGE. 
%(the command "\vskip 4 truecm" actually results in about 4.5 cm of empty
%space at the top, or about 19.6%).  The publisher will probably reset the 
%chapter heading (your title and by-line), but you should follow my 
%19-20% prescription anyway!  In my design I am following the Les Houches 
%lecture notes volume produced by Nova.   If you have LOTS of authors and 
%by-lines you may want to allow a bit less space at the top (e.g. if you 
%have 3 or more sets of authors and institutions).
\topinsert
\vskip 3.2 truecm
\endinsert
\centerline{\bigbfont NOVEL EXCITONIC STATES AND PHOTOLUMINESCENCE}
%If your title is only one line long, put a % before the 2nd title line.
%If your title is longer than two lines, continue thus:
\vskip 6 truept
\centerline{\bigbfont IN QUANTUM HALL SYSTEMS}
%Don't forget to remove the % sign from the preceding line if you use it! 
\vskip 20 truept
%Now comes your by-line with institutional addresses.  
\centerline{\bigifont J. J. Quinn and A. Wojs$^\dagger$}
\vskip 8 truept
\centerline{\bigrfont Department of Physics, University of Tennessee}
\vskip 2 truept
\centerline{\bigrfont Knoxville, Tennessee 37996, USA}
\vskip 1.8 truecm

\centerline{\bf 1.  INTRODUCTION}
\vskip 12 truept

Photoluminescence (PL) in quantum Hall systems has been studied for
more than a decade [1--3].
The possibility of extracting from the experimental data information 
about correlations within the underlying two-dimensional electron gas
(2DEG) in a high magnetic field has stimulated their effort.
For the ideal theoretical model, in which electrons and holes lie 
on the same defect free 2D layer at extremely high magnetic field, 
the PL spectrum is predicted to consist of a single line, resulting 
from radiative decay of a neutral exciton ($X$), and to give no 
information about electron-electron correlations [4].
This results from a ``hidden symmetry'' associated with the equality 
of magnitudes of the particle-particle interactions, 
$V_{ee}=V_{hh}=-V_{eh}$.

In realistic systems, the admixture of higher Landau levels (LL's) 
by the Coulomb interaction, the alteration of the interactions by 
form factors associated with finite quantum well width, and the 
finite separation of the electron and hole layers by a built-in 
or an applied electric field lead to a much richer spectrum.
Both neutral ($X$) and charged ($X^-$) excitons contribute to PL in 
fractional quantum Hall systems at filling factors $\nu\le{1\over3}$
[3].
In the ideal model only the triplet $X^-_{\rm td}$ (with electronic
spin $S=1$) is bound and it is non-radiative (subscript ``d'' denotes 
dark).
In real experimental systems, both the singlet, $X^-_{\rm s}$, 
and triplet, $X^-_{\rm t}$, have one or more bound states.
The PL intensity of the $X^-_{\rm td}$ ground state is non-zero,
but it is weak compared to the $X^-_{\rm s}$ and $X$.
In fact, it was observed only recently [3], when special care 
(e.g., very low temperature) was taken to detect its weak PL signal.
In order to understand the PL process in real systems, we study 
small systems by numerical diagonalization (in spherical geometry).
For the simplest systems, the Hilbert space is not restricted to the 
lowest LL as in the ideal theoretical model, but linear combinations 
containing up to five LL's are employed.
The Coulomb interaction is modified so as to mimic that of a 
symmetric quantum well of width $w=11.5$~nm.
For systems containing more than three electrons and one hole, 
we fall back to the single LL approximation, but we allow the 
electrons and holes to reside on distinct 2D layers separated by 
a distance $d$.
This breaks the ``hidden symmetry'' in the simplest way, but it 
is a reasonable approximation only for the highest magnetic fields.

The numerical studies [5] at $\nu\le{1\over3}$ were done for 
electron systems which are maximally spin polarized (except
for the spin reversed electron of the $X^-_{\rm s}$).
It is known that for $\nu=1$ the ground state is spin polarized, 
and the lowest energy excitations are spin waves (SW's).
However, when $\nu$ is slightly smaller than unity, a ``spin hole,'' 
$h$, in the $\nu=1$ level is present, and it can spontaneously
create and bind a SW [6].
Depending on the Zeeman energy, $E_{\rm Z}$, the ground state will 
contain either free spin holes or ``antiskyrmions,'' $S^+$ (bound 
states of one or more SW's and the $h$).
For $\nu$ slightly larger than unity, the reversed spin electrons,
$e_{\rm R}$ (above the $\nu=1$ filled level) can form skyrmions,
$S^-$, states containing one or more SW's bound to $e_{\rm R}$.
We review some results on spin excitations within the ideal model 
[7], with the thought of simply replacing on of the spin holes of the 
$\nu=1$ states by a valence band hole, $v$.
In this ideal model the spin hole, $h$, and the valence band hole, 
$v$, are distinguishable (they can be distinguished by a pseudospin), 
but their interactions with other charged particles are equal in 
magnitude.
This implies that an $X^+=(vhe_{\rm R})^+$ made up of a $v$ bound 
to a SW must exist.
We consider PL resulting from this antiskyrmion-like excitonic 
complex.
For $\nu$ slightly smaller than unity, some antiskyrmions, $S^+$,
are present before introduction of the $v$.
These will avoid the $X^+$ due to Coulomb repulsion, so the PL
at $1-\nu\ll1$ should be due to isolated $X^+$ complexes
(and $X$'s [8]).
For $\nu$ slightly larger than unity, some $S^-$ complexes will
be present before introduction of the $v$.
It is possible that the $v$ captures $e_{\rm R}$'s  from these 
$S^-$ complexes, that is $v^++S^-\rightarrow X+{\rm SW}$ or 
$v^++2S^-\rightarrow X^-+2\;{\rm SW}$ or that the $v$ creates 
and binds a SW to form a $(vhe_{\rm R})^+=X^+$.
The $X^+$ has an attractive interaction with the skyrmions that
could result in $X^++S^-\rightarrow X^-+h^++{\rm SW}$, or it could
decay radiatively before reaching the $(X^-,h^+)$ state, particularly
when the initial $S^-$ density is very small.
From analogy with the dark triplet $X^-_{\rm td}$ at $\nu\le{1\over3}$,
we can guess that isolated $X^-$ will be ``dark,'' and that the 
$v$-$e$ recombination will be primarily with majority spin electrons
as suggested by Cooper and Chklovskii [8].
Definitive predictions will require numerical work based on more 
realistic  models than used in much of their work.
This makes the present work qualitative and suggestive, and not 
intended for detailed comparison with experiment.
\vskip 28 truept

\centerline{\bf 2.  ENERGY SPECTRUM AND PL FOR $\nu\leq{1\over3}$}
\vskip 12 truept

It has become rather standard to diagonalize numerically the Coulomb
interaction for a finite system of $N$ electrons confined to
a spherical surface which contains at its center a magnetic monopole
of strength $2Q$ flux quanta [9].
In the ideal theoretical model only states of the lowest LL are included.
For realistic experimental systems (having a finite quantum well
width $w$ in a finite magnetic field $B$) both higher LL's and the 
modification for the Coulomb matrix elements associated with the 
envelope functions of the quantum well must be included.

\topinsert
\input psfig.sty
\centerline{\hskip10mm\psfig{figure=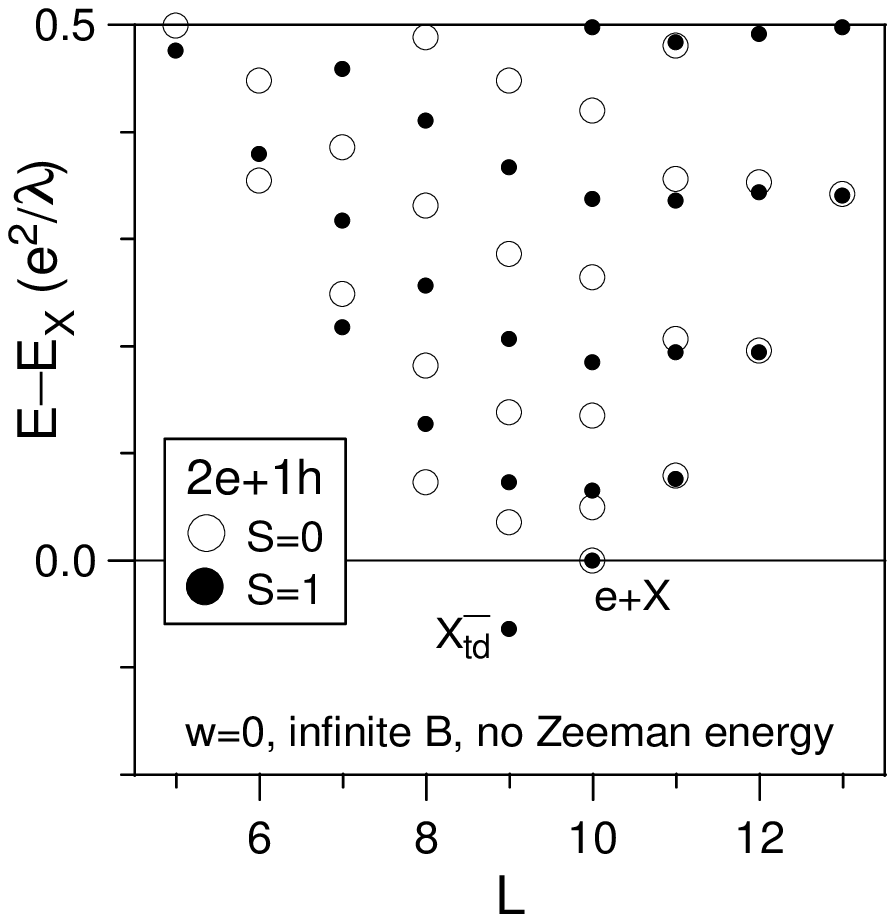,width=11truecm}}
\vskip -0.1truecm 
\noindent
{\bf Figure 1.} 
Energy spectrum (energy $E$ vs.\ angular momentum $L$) 
of the $2e+v$ system in the lowest LL, calculated on a Haldane 
sphere for monopole strength $2Q=20$.
$E_X$ is the exciton energy, and $\lambda$ is the magnetic length.
\vskip 12truept
\endinsert
In Fig.~1 we present the energy spectrum for a simple $2e+v$ system
at $2Q=20$ evaluated in the ideal theoretical model and excluding 
$E_{\rm Z}$ [5].
The solid dots are triplet states ($S=1$), and the open circles 
are singlets ($S=0$).
The state labeled $e+X$ at angular momentum $L=10$ is a 
``multiplicative state'' consisting of an unbound electron and 
an $X$.
Notice that only one bound state ($X^-_{\rm td}$) occurs.
It is at $L=9$ and is called the ``dark triplet'' because it is 
forbidden to decay radiatively.

\topinsert
\input psfig.sty
\centerline{\hskip10mm\psfig{figure=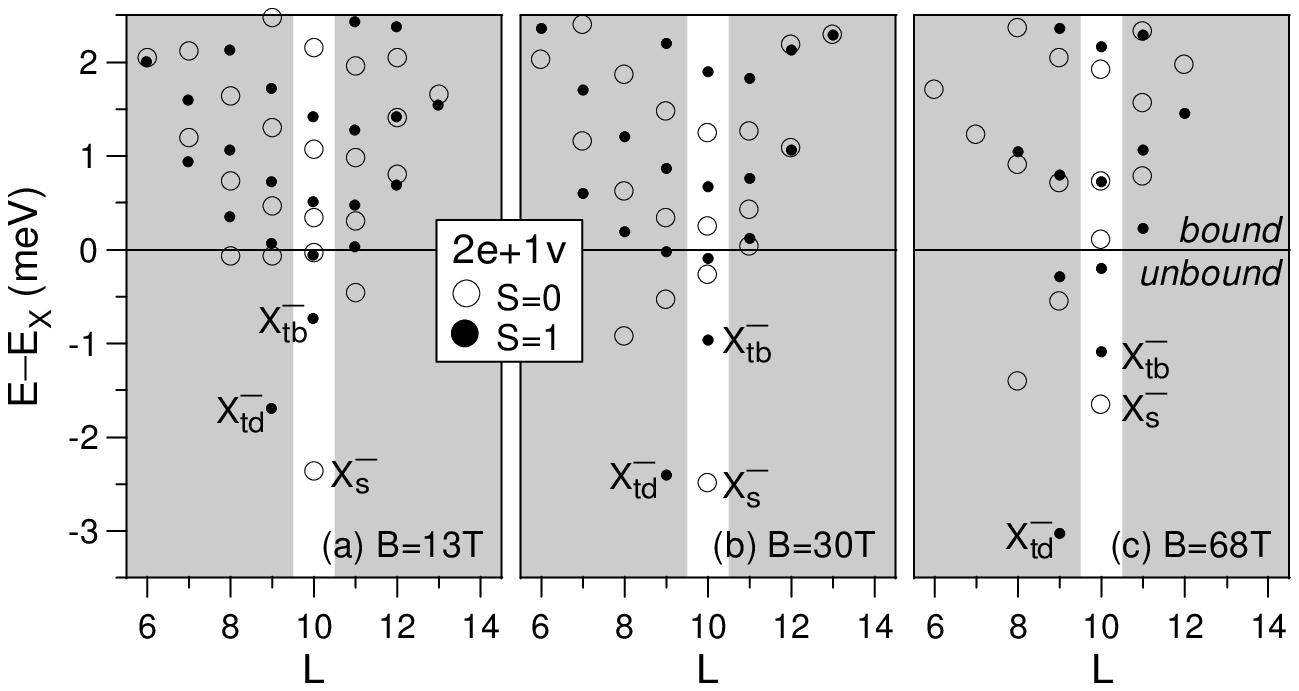,width=11truecm}}
\vskip 0.0truecm 
\noindent
{\bf Figure 2.} 
Same as Fig.~1, but for a realistic GaAs well of width $w=11.5$~nm 
at the finite values of magnetic field $B$ as shown.
The Zeeman energy has been included, and five LL's for both electrons 
and hole have been used in the diagonalization.
\vskip 12truept
\endinsert
In Fig.~2 similar results are presented for a realistic system
consisting of a symmetric GaAs quantum well of width $w=11.5$~nm at
the finite values of the magnetic field $B=13$, 30, and 68~T.
The appropriate electron Zeeman splitting has been included.
To achieve even qualitative agreement with experimental data,
it has also been necessary to include a number of higher LL's, 
particularly at the lower magnetic fields.
Five LL's were needed to obtain convergence in our calculations.

In Fig.~2c, at the high magnetic field of 68~T, the $X^-_{\rm td}$ 
at $L=9$ is still the lowest energy state, but singlet and another 
triplet bound states occur at $L=8$, 9, and 10.
The singlet at $L=8$ (no label) and $L=10$ ($X^-_{\rm s}$), and 
the triplet at $L=10$ ($X^-_{\rm tb}$) have roughly half the 
binding energy of the $X^-_{\rm td}$ ground state.
As the magnetic field is lowered, the $X^-_{\rm s}$ at $L=10$ 
moves down in energy relative to the triplet states, becoming the 
ground state for $B$ smaller than 30~T (as shown in Figs.~2b and a). 
This is in agreement with the results obtained by Whittaker and 
Shields [10].
The spectra are quite sensitive to the experimental parameters.
The well width $w$ enters the Coulomb interaction [5] through 
$V(r)=e^2/\sqrt{r^2+d^2}$, where $d$ is proportional to $w$.
The cyclotron frequencies $\omega_{ce}(B)$ and $\omega_{cv}(B)$ 
for the electrons and valence hole, and the Zeeman energy, 
$E_{\rm Z}(B)$, are taken from experiment, after Refs.~[11,12].

Because exact diagonalization gives the eigenfunctions as well as 
the eigenvalues, it is straightforward to evaluate matrix elements 
of the PL operator $\hat L=\int\,{\rm d}^2r\,\hat\Psi_e(r)\hat
\Psi_v(r)$ between an initial $Ne+v$ state $\Phi_i$ and final 
$(N-1)e$ states $\Phi_f$.
$\hat\Psi_e$ and $\hat\Psi_v$ are the annihilation operators for 
an electron and valence hole respectively.
The transition oscillator strength [13] is proportional to 
$|\left<\right.\!\Phi_f|\,\hat L\,|\Phi_i\!\left.\right>|^2$.
For an isolated $X^-$ (where $N=2$), angular momentum conservation 
forbids the lowest triplet ($X^-_{\rm td}$) from decaying radiatively.
The singlet $X^-_{\rm s}$ and the excited triplet $X^-_{\rm tb}$ 
(``b'' stands for ``bright'') both have finite (and comparable) 
oscillator strengths.
These radiative states appear in the light color area at $L=10$ in 
Fig.~2.

When additional electrons are present ($N>2$) radiative decay of the
$X^-_{\rm td}$ is not strictly forbidden, since in the recombination 
process an unbound electron can scatter, changing the momentum of the 
final state.
However, it was found that for $\nu\leq{1\over3}$ such decays are weak
because Laughlin correlations of the $X^-$ with unbound electrons
inhibit close collisions.
The amplitude for radiative decay of the $X^-_{\rm td}$ is estimated 
[5] to be smaller by one or more orders of magnitude than those of 
the $X^-_{\rm s}$ and $X^-_{\rm tb}$.
It was suggested in [5] that the $X^-_{\rm td}$ would be difficult 
to see in PL, and that the non-crossing peaks observed by Hayne 
{\sl et al.} [14] were the $X^-_{\rm s}$ and $X^-_{\rm tb}$.
The presence of impurities relaxes the $\Delta L=0$ selection rule,
and the $X^-_{\rm td}$ peak is clearly observed at very low temperature
where the excited $X^-_{\rm tb}$ and $X^-_{\rm s}$ states are sparsely 
populated [3].
The agreement of experiment [3] and the numerical predictions [5] 
reinforce the hope of using PL to understand correlations in quantum 
Hall systems.
\vskip 28 truept

\centerline{\bf 3.  FRACTIONALLY CHARGED EXCITONS}
\vskip 12 truept

For systems containing more than two or three electrons and 
a valence hole, it is difficult to include the admixture 
of higher LL's caused by Coulomb interactions.
However, the ``hidden symmetry'' can be broken by separating
the $e$ and $v$ planes by a finite length $d$.
Because only the lowest LL is included, such a simple model 
will be useful only at the highest magnetic fields.
When $d$ is measured in units of magnetic length $\lambda=
\sqrt{\hbar c/eB}$, we can identify three regimes: strong 
($e$--$v$) coupling when $d\ll1$; weak coupling when $d\gg1$, 
and intermediate coupling when $d\sim1$.

In the strong coupling region, neutral ($X$) and charged triplet 
excitons ($X^-$) are formed due to the strong $e$-$v$ interactions.
Neutral excitons in their $L=0$ ground state are almost decoupled
from the remaining $N-1$ electron system.
At $d=0$, the hidden symmetry holds, and the excitons are 
completely decoupled, these states are called multiplicative 
states.
The $X^-$ is a charged Fermion that interacts with the remaining
$N-2$ electrons.
The generalized composite Fermion (CF) picture [15] describes the 
low lying excitations and the Laughlin correlations very well.

For $d\gg1$, the $e$-$v$ interaction is a weak perturbation on the
energies of the $N$ electron system.
The low energy spectrum contains bands obtained by adding the 
angular momentum of the hole ($l_v=Q$) to that of the low lying
quasiparticle states of the electrons, obtained from the
CF picture or from numerical diagonalization [13].

For intermediate coupling, the $e$-$v$ interaction is not strong
enough to bind a full electron, but bound states of one or more
Laughlin quasielectrons (QE) do occur.
These are referred to as fractionally charged excitons (FCX)
or anyon excitons.

\topinsert
\input psfig.sty
\centerline{\hskip10mm\psfig{figure=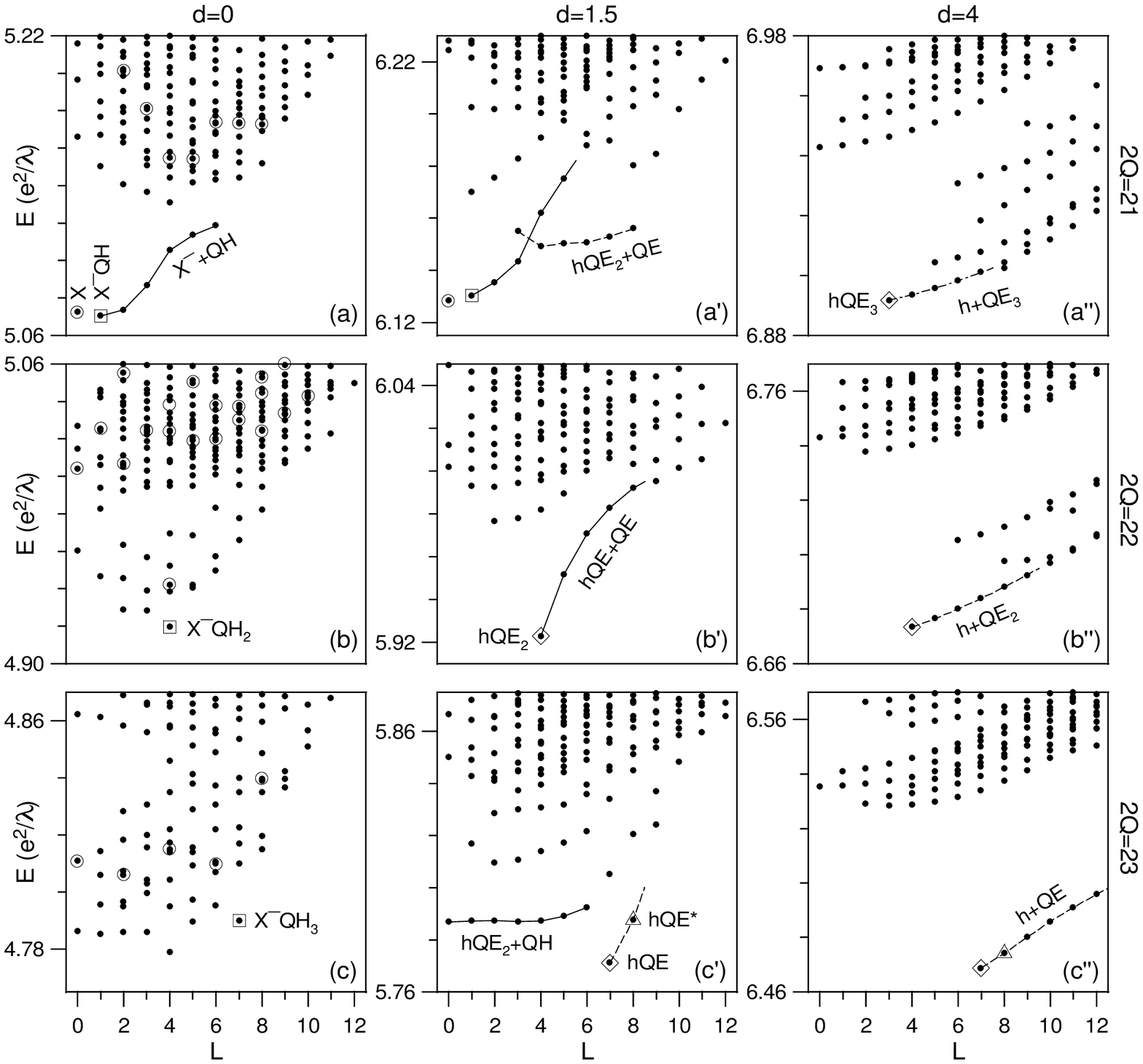,width=11truecm}}
\vskip -0.2truecm 
\noindent
{\bf Figure 3.} 
Energy spectra (energy $E$ vs.\ angular momentum $L$) 
of the $9e+v$ system in the lowest LL, calculated on 
a Haldane sphere for different monopole strengths $2Q=21$,
22, and 23, and $e$-$v$ layer separations $d=0$, 1.5, and 4.
Lines and open symbols mark the low-energy states 
containing different bound excitonic complexes.
\vskip 12truept
\endinsert
In Fig.~3 we display the energy spectra for a $9e+v$ system at 
$2Q=21$, 22, and 23, for $d=0$, 1.5, and 4.
This data is taken from Ref.~[16].
It is worth noting that for $d=0$, the $X$ and the $X^-$ appear
at each value of $2Q$, while for $d=4$ only bound states of the
valence hole and Laughlin quasielectrons (QE) occur.
In frame (a) the low lying state at $L=0$ is a multiplicative 
state consisting of one decoupled neutral $X$ and a Laughlin
condensed state of the remaining eight electrons [$2Q=3(N-1)=21$
for this Laughlin $\nu={1\over3}$ state].
A band of states extending from $L=1$ to $L=6$ consists of an
$X^-$ with $l_{X^-}={5\over2}$ and a single Laughlin quasihole
(QH) with $l_{\rm QH}={7\over2}$.
The solid line drawn through these states represents the 
pseudopotential $V_{X^-,{\rm QH}}(L)$ of the interacting
$X^-$--QH pair.
Initially, this band of states was incorrectly interpreted as 
a neutral exciton interacting with a magneto-roton [17].
The generalized CF picture [15] gives a simple and natural
interpretation of all of the low lying states for all values
of the parameter $d$.
In frame (b'') the nine electron system contains two QE's.
Since $l_{\rm QH}=Q-(N-1)=3$ and $l_{\rm QE}=l_{\rm QH}+1$,
therefore the lowest CF shell is filled by $2l_{\rm QH}+1=7$
of the CF's leaving two as QE's in the first excited CF shell.
The allowed values of the total angular momentum of their pair
are $L_{\rm 2QE}=1\oplus3\oplus5\oplus7$.
In the mean field CF picture, these pair states would form
a degenerate band.
Exact diagonalization (which includes QE-QE interactions 
beyond mean field) gives the ordering of the energies 
$E(L_{\rm 2QE})$ as $E(7)<E(3)<E(5)<E(1)$.
The interaction of the valence hole (with angular momentum
$l_h=Q=11$) with the QE pair leads to four bands, each 
increasing with $\vec{L}=\vec{L}_{\rm 2QE}+\vec{l}_h$.
Detailed discussion of all these spectra is given in Ref.~[16].
In this review we restrict the discussion to a few 
illustrative examples.
\vskip 28 truept

\centerline{\bf 4.  SPIN EXCITATIONS NEAR $\nu=1$}
\vskip 12 truept

In order to understand the excitonic complexes that can be formed
near filling factor $\nu=1$, it is first necessary to study the 
elementary excitations that can occur in the absence of valence 
band holes.
For $\nu=1$, the lowest energy excitations are spin flip excitations 
which create an $e_{\rm R}$ in the same $n=0$ LL leaving behind an 
$h$ in the otherwise filled $\nu=1$ state.
Even when $E_{\rm Z}=0$, the Coulomb exchange energy will 
spontaneously break the spin ($\uparrow,\downarrow$) symmetry 
giving a polarized ground state.

\topinsert
\input psfig.sty
\centerline{\hskip10mm\psfig{figure=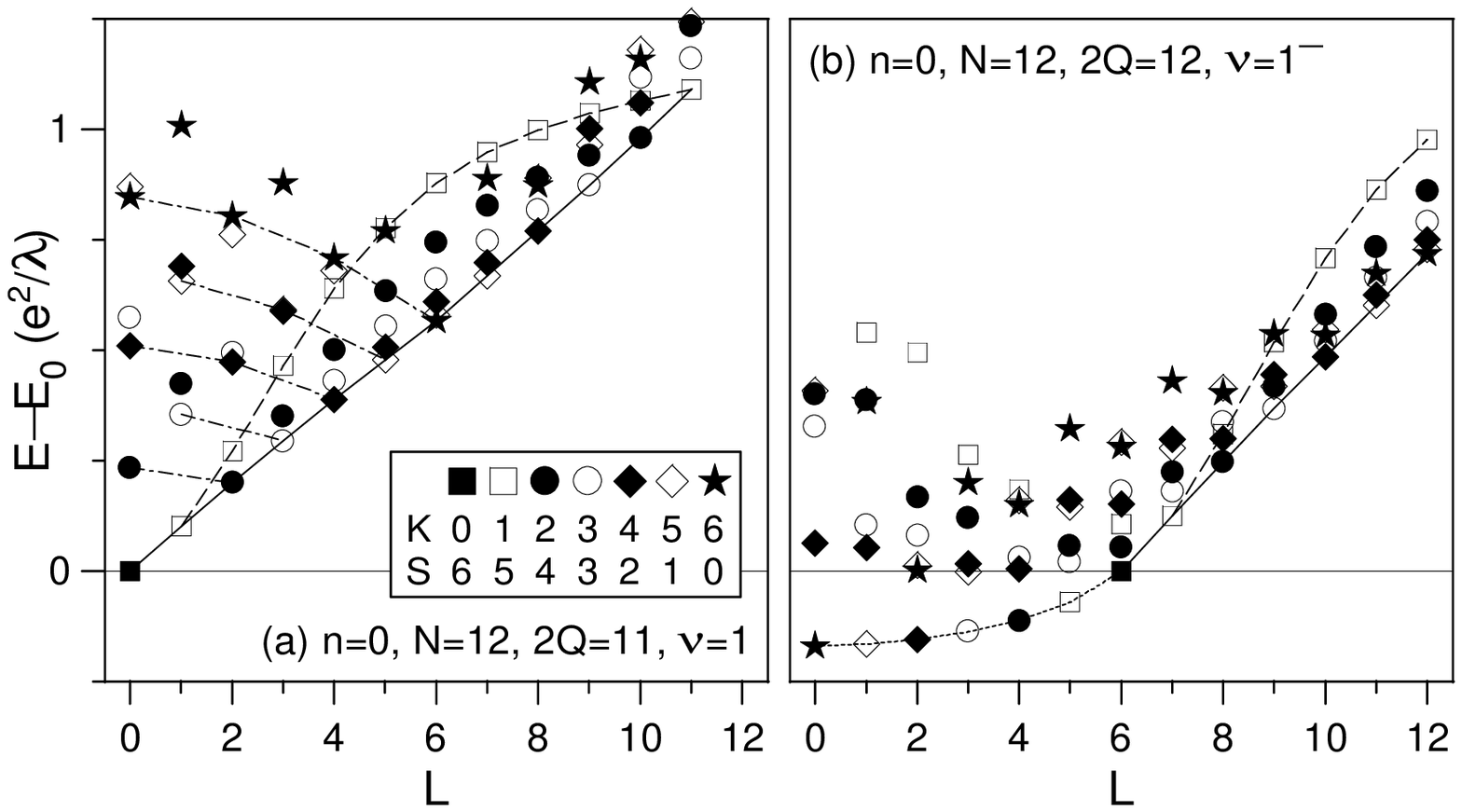,width=11truecm}}
\vskip -0.1truecm 
\noindent
{\bf Figure 4.} 
Energy spectra (energy $E$ vs.\ angular momentum $L$) 
of the spin unpolarized $12e$ system in the lowest LL, 
calculated on a Haldane sphere for monopole strength 
$2Q=11$ (a) and 12 (b).
\vskip 12truept
\endinsert
In Fig.~4a we show the low lying spin excitations of the $\nu=1$ state 
(with $E_{\rm Z}$ taken to be zero) for a system of $N=12$ electrons [7].
The solid square at $L=0$ is the spin polarized $\nu=1$ ground state 
with spin $S=6$.
The symbol $K={1\over2}N-S$ is the number of spin flips away from
the fully spin polarized state.
The band of open squares connected by a dashed line gives the SW
dispersion $\varepsilon_{\rm SW}(L)$.
The angular momentum $L$ is related to wave number $k$ by $L=kR$,
where $R$ is the radius of the sphere.
confined.
The SW consists of a single $e_{\rm R}h$ pair; its dispersion can 
be evaluated analytically [18].
The solid circles, open circles, etc. represent states containing
2, 3, \dots\ spin flips (i.e., 2, 3, \dots\ $e_{\rm R}h$ pairs).
Dot-dashed lines connect low lying states with equal numbers of spin
flips.
It is interesting to note the almost straight line connecting the 
lowest energy states at $0\leq L\leq6$.
This can be interpreted as band of $K$ SW's each with $l_{\rm SW}=1$ 
with $L=K$.
The near linearity suggests that these $K$ SW's are very nearly
non-interacting in the state with $L=K$.

In Fig.~4b we present the low energy spectrum for $\nu=1^-$ (i.e., 
a single spin hole in the $\nu=1$ state).
In both Fig.~4a and b only the lowest energy states at each $L$ and 
$S$ are shown.
Of particular interest in Fig.~4b is the band of states with $L=S=Q-K$ 
and negative energy.
These are antiskyrmion states, $S_K^+=Ke_{\rm R}+(K+1)h$, bound states
of one $h$ and $K$ SW's [6,7].
They are analogous to interband charged excitons [5], but they can be 
equilibrium states not subject to radiative decay at the appropriate 
value of $E_{\rm Z}$.
Skyrmion states are $S_K^-=Kh+(K+1)e_{\rm R}$.
Electron-hole symmetry requires their existence for $\nu>1$.

It has been demonstrated [7] that in the fractional quantum Hall 
regime analogous excitations occur with QE$_{\rm R}$ and QH replacing 
$e_{\rm R}$ and $h$ of the integral quantum Hall case.
SW's, skyrmions, and antiskyrmions made from Laughlin 
quasiparticles occur for $\nu\approx{1\over3}$.
The most stable skyrmion or antiskyrmion size depends weakly on the 
quantum well width for the $\nu\approx1$ state, but for $\nu\approx3$, 
5, \dots\ the well width $w$ must be of the order of a few times the 
magnetic length in order to obtain stable bound states of SW's and 
$h$'s or $e_{\rm R}$'s [7,19].
As reported in Ref.~[20], the admixture of higher LL's caused by 
the Coulomb interaction weakly affects the skyrmion energy spectrum, 
particularly when the finite well width $w$ is also taken into account.

The skyrmion and antiskyrmion states $S_K^\pm$ are quite analogous 
to the excitonic $X_K^\pm$ states of valence band holes interacting 
with conduction band electrons.
In the ideal theoretical model, a valence hole has exactly the same 
interactions as a spin hole in the $\nu=1$ state of the conduction 
band.
In fact these two types of holes can probably be distinguished by an
pseudospin as is done for electrons on different layers of a bilayer
system [21].
The spectrum and possible condensed states of a multicomponent Fermion
liquid containing electrons, $X_1^-$, $X_2^-$, etc., has been
considered in Ref.~[15].
Exactly the same ideas are applicable to a liquid of electrons and 
skyrmions or antiskyrmions of different sizes.
The only difference is that $S^-=h(e_{\rm R})_2$ is stable while 
the $X^-=ve_2$ has a finite lifetime for radiative $e$-$v$ 
recombination.

When there are $N_h$ spin holes in the $\nu=1$ level (or $N_e$ reversed 
spin electrons in addition to the filled $\nu=1$ level) and when $N_h$
(or $N_e$) is much smaller than $N\approx2Q+1$, the degeneracy of the 
filled lowest LL, then the most stable configuration will consist of 
$N_h$ antiskyrmions (or $N_e$ skyrmions) of the most stable size.
These antiskyrmions (or skyrmions) repel one another.
They are positively (or negatively) charged Fermions with standard LL
structure, so it is not surprising that they would form either a Wigner
lattice or a Laughlin condensed state with $\nu$ for the antiskyrmion
(or skyrmion) equal to an odd denominator fraction as discussed in 
Refs.~[7,22,23].
\vskip 28 truept

\centerline{\bf 5.  PHOTOLUMINESCENCE NEAR $\nu=1$}
\vskip 12 truept

In the ideal theoretical model, a valence hole acts exactly like 
a spin hole in the $\nu=1$ level of the conduction band.
Therefore we would expect an excitonic complex consisting of $K$ SW's
bound to the valence hole to be the lowest energy state, in the same 
way that the antiskyrmion consisting of $K$ SW's bound to a spin 
hole in the $\nu=1$ level gives the lowest energy state when $E_{\rm Z}$
is sufficiently small.
For a small number of valence holes, the $X_K^+=v(e_{\rm R}h)_K$ 
excitonic complexes formed by each valence hole will repel one another.
If a small number of antiskyrmions are already present (for $\nu<1$), 
the $S^+$-$X^+$ repulsion will lead to Laughlin correlations or Wigner 
crystallization of the multicomponent Fermion liquid.
Just as for the $X^-$'s in the dilute regime, the PL at low temperature 
will be dominated by the $X_K^+\rightarrow S_{K'}^++\gamma$ process, 
with $K'=K$ or $K-1$ depending on spin of the annihilated $v$ (i.e., 
on the circular polarization of the emitted photon $\gamma$).
This corresponds to the most stable $X_K^+$ undergoing radiative 
$e$-$v$ or $e_{\rm R}$-$v$ recombination and leaving behind an 
antiskyrmion consisting of $K$ or $K-1$ SW's bound to a spin hole 
of the $\nu=1$ state.
Because the valence hole and the spin hole in the $\nu=1$ conduction 
level are distinguishable (or have different pseudospin) even in the 
ideal theoretical model this PL is not forbidden.
It will be very interesting to see how realistic sample effects 
(finite well width, LL admixture, finite separation between the 
electron and valence hole layers) alter the conclusions of the 
ideal theoretical model.

For $\nu\geq1$, negatively charged skyrmions are present before the
introduction of the valence holes.
The skyrmions are attracted by the $X_K^+$ charged exciton, but 
how this interaction affects the PL can only be guessed.
It is possible that the interaction of the valence hole with the 
skyrmions will lead to the formation of an $X$ or an $X^-_{\rm td}$
and SW's.
The $X^-_{\rm td}$ will be very weakly radiative (just as in the case
of $\nu\le{1\over3}$).
However, the recombination can occur with a majority spin electron.
This case was considered in Ref.~[8] for the case of a single 
$X^-_{\rm td}$.
The largest oscillator strength occurred for the process $X^-_{\rm td}
\rightarrow S_1^-+\gamma$; in other words the spin hole left in the
$\nu=1$ level  by the $ve$ recombination formed a bound state with 
the two reversed spin electrons of the $X^-_{\rm td}$.
Lower energy photon resulted when the two reversed spin electrons
and the spin hole did not form the $S_1^-$ bound state.

\topinsert
\input psfig.sty
\centerline{\hskip10mm\psfig{figure=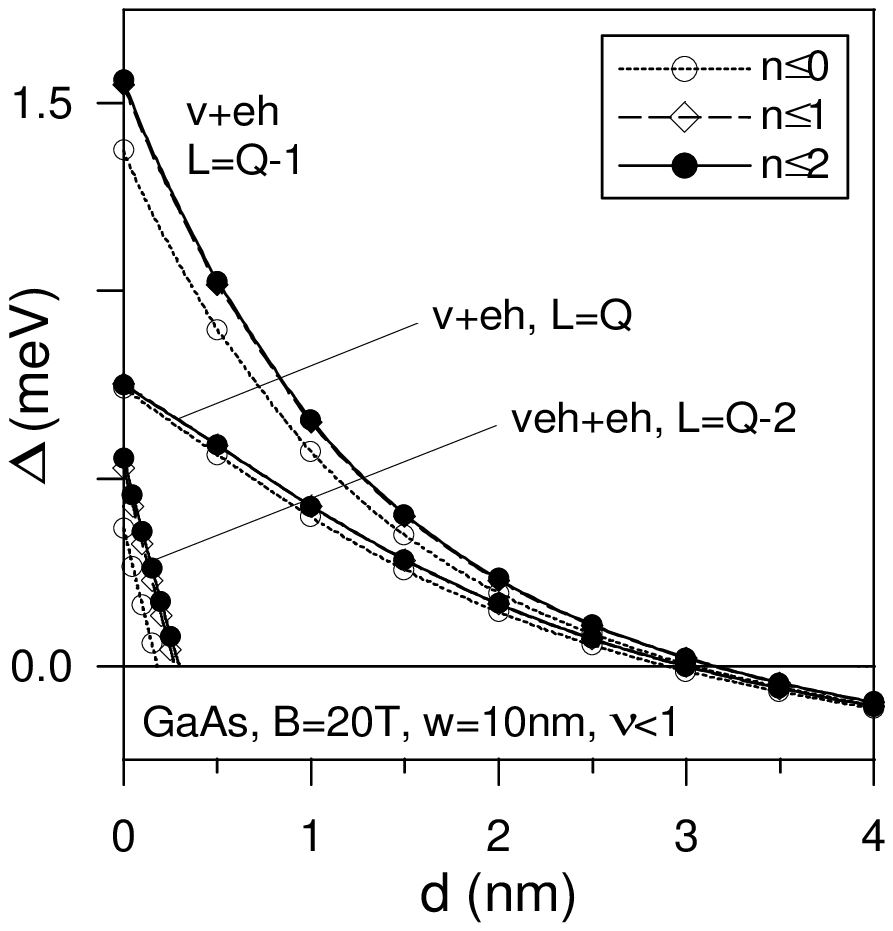,width=11truecm}}
\vskip 0.0truecm 
\noindent
{\bf Figure 5.} 
Binding energies $\Delta$ of charged skyrmion exciton 
$X_1^+=v(e_{\rm R}h)$ at angular momentum $L=Q-1$ and $Q$, 
and of charged skyrmion biexciton $X_2^+=v(e_{\rm R}h)_2$ 
at $L=Q-2$ calculated on a sphere for monopole strength 
$2Q=30$ and plotted as a function of the $e$-$v$ layer 
separation $d$.
The parameters are for a GaAs well of width $w=10$~nm at 
a magnetic field $B=20$~T.
Different curves include 1--3 LL's for the $v$.
\vskip 12truept
\endinsert
We are currently investigating more realistic models in systems 
containing a small number of $S^\pm$'s and $X^\pm$'s.
As a preliminary example of our results we show in Fig.~5 the 
binding energy, $\Delta$, of the $X_1^+=v(e_{\rm R}h)$ and $X_2^+=
v(e_{\rm R}h)_2$ complexes for different values of the total angular 
momentum $L$ as a function of the $e$-$v$ layer separation $d$.
$\Delta$ is defined as the binding energy of the SW to a charged 
complex, $\Delta[v(e_{\rm R}h)]=E[v]+E[e_{\rm R}h]-E[v(e_{\rm R}h)]$ 
and $\Delta[v(e_{\rm R}h)_2]=E[v(e_{\rm R}h)]+E[e_{\rm R}h]-
E[v(e_{\rm R}h)_2]$.
The calculation was done for a GaAs quantum well of width $w=10$~nm, 
at a magnetic field of 20~T, but $E_{\rm Z}$ has been omitted.
The different symbols (open circles, open diamonds, and solid 
circles) are for calculations in which one, two, or three LL's 
for the valence hole have been included (inter-LL excitations 
of conduction electrons are less important due to their smaller 
effective mass).
It is clear that the binding energies decrease with increasing 
$e$-$v$ layer separation as expected.
\vskip 28 truept

\centerline{\bf ACKNOWLEDGMENTS}
\vskip 12 truept

The authors wish to acknowledge partial support from the Materials 
Research Program of Basic Energy Sciences, US Department of Energy
and thank I.~Szlufarska for helpful discussions.
JJQ thanks the US Army Research Office--Research Triangle Park 
for travel support through a grant to Southern Illinois 
University--Carbondale.  
\vskip 28 truept

\centerline{\bf REFERENCES}
\vskip 12 truept

\item{$\dagger$}
On leave from: 
Institute of Physics, Wroclaw University of Technology (Poland).

\item{[1]}
D.~Heiman, B.~B.~Goldberg, A.~Pinczuk, C.~Tu, A.~C.~Gossard, 
and J.~H.~English,
{\it Phys.~Rev.~Lett.} {\bf61}, 605 (1988);
A.~J.~Turberfield, S.~R.~Haynes, P.~A.~Wright, R.~A.~Ford, 
R.~G.~Clark, J.~F.~Ryan, J.~J.~Harris, and C.~T.~Foxon,
{\it Phys.~Rev. Lett.} {\bf65}, 637 (1990);
B.~B.~Goldberg, D.~Heiman, A.~Pinczuk, L.~N.~Pfeiffer, and K.~West,
{\it Phys.~Rev.~Lett.} {\bf65}, 641 (1990).

\item{[2]}
Z.~X.~Jiang, B.~.D.~McCombe, and P.~Hawrylak,
{\it Phys.~Rev.~Lett.} {\bf 81}, 3499 (1998);
H.~A.~Nickel, G.~S.~Herold, T.~Yeo, G.~Kioseoglou, Z.~X.~Jiang, 
B.~D.~McCombe, A.~Petrou, D.~Broido, and W.~Schaff,
{\it Phys.~Status Solidi B} {\bf 210}, 341 (1998);
L.~Gravier, M.~Potemski, P.~Hawrylak, and B.~Etienne,
{\it Phys.~Rev.~Lett.} {\bf80}, 3344 (1998);
H.~Buhmann, L.~Mansouri, J.~Wang, P.~H.~Beton, N.~Mori, 
M.~Heini, and M.~Potemski,
{\it Phys.~Rev.~B} {\bf51}, 7969 (1995);
T.~Wojtowicz, M.~Kutrowski, G.~Karczewski, J.~Kossut, 
F.~J.~Teran, and M.~Potemski,
{\it Phys.~Rev.~B} {\bf59}, R10 437 (1999).

\item{[3]}
G.~Yusa, H.~Shtrikman, and I.~Bar-Joseph,
{\it Phys.~Rev.~Lett.} {\bf 87}, 216402 (2001);
T.~Vanhoucke, M.~Hayne, M.~Henini, and V.~V.~Moshchalkov,
{\it Phys.~Rev.~B} {\bf63}, 125331 (2001);
C.~Schuller, K.~B.~Broocks, C.~Heyn, and D.~Heitmann,
{\it Phys.~Rev.~B} {\bf65}, 081301 (2002).

\item{[4]}
A.~H.~MacDonald and E.~H.~Rezayi, 
{\it Phys.~Rev.~B} {\bf 42}, 3224 (1990);
A.~B.~Dzyu\-benko and Yu.~E.~Lozovik,
{\it J.~Phys.~A} {\bf 24}, 414 (1991).

\item{[5]}
A.~W\'ojs, J.~J.~Quinn, and P.~Hawrylak,
{\it Phys.~Rev.~B} {\bf 62}, 4630 (2000);
I.~Szlufarska, A.~W\'ojs, and J.~J.~Quinn,
{\it Phys.~Rev.~B} {\bf 63}, 085305 (2001).

\item{[6]}
S.~L.~Sondhi, A.~Karlhede, S.~A.~Kivelson, and E.~H.~Rezayi,
{\it Phys.~Rev.~B} {\bf 47}, 16419 (1993);
H.~A.~Fertig, L.~Brey, R.~C\^ot\'e, and A.~H.~MacDonald,
{\it Phys.~Rev. B} {\bf 50}, 11018 (1994);
A.~H.~MacDonald, H.~A.~Fertig, and L.~Brey,
{\it Phys.~Rev.~Lett.} {\bf 76}, 2153 (1996);
H.~A.~Fertig, L.~Brey, R.~C\^ot\'e, and A.~H.~MacDonald,
{\it Phys.~Rev. Lett.} {\bf 77}, 1572 (1996);
X.~C.~Xie and S.~He,
{\it Phys.~Rev.~B} {\bf53}, 1046 (1996);
H.~A.~Fertig, L.~Brey, R.~C\^ot\'e, A.~H.~MacDonald, A.~Karlhede,
and S.~L.~Sondhi,
{\it Phys.~Rev.~B} {\bf 55}, 10671 (1997).

\item{[7]}
A.~W\'ojs and J.~J.~Quinn, 
{\it Solid.~State.~Commun.} {\bf 122}, 407 (2002);
A.~W\'ojs and J.~J.~Quinn, 
{\it Phys.~Rev.~B} {\bf 66}, 045323 (2002).

\item{[8]}
N.~R.~Cooper and D.~B.~Chklovskii,
{\it Phys.~Rev.~B} {\bf55}, 2436 (1997);
T.~Portengen, J.~R.~Chapman, V.~N.~Nicopoulos, 
and N.~F.~Johnson,
{\it Phys.~Rev.~B} {\bf56}, R10052 (1997).

\item{[9]}
F.~D.~M.~Haldane,
{\it Phys.~Rev.~Lett.} {\bf 51}, 605 (1983);
F.~D.~M.~Haldane and E.~H.~Rezayi,
{\it Phys.~Rev.~Lett.} {\bf 60}, 956 (1988).

\item{[10]}
D.~M.~Whittaker and A.~J.~Shields,
{\it Phys.~Rev.~B} {\bf 56}, 15185 (1997).

\item{[11]}
B.~E.~Cole, J.~M.~Chamberlain, M.~Henini, T.~Cheng, W.~Batty, 
A.~Wittlin J.~A.~A.~J.~Perenboom, A.~Ardavan, A.~Polisski, 
and J.~Singleton,
{\it Phys.~Rev.~B} {\bf 55}, 2503 (1997).

\item{[12]}
M.~J.~Snelling, G.~P.~Flinn, A.~S.~Plaut, R.~T.~Harley, 
A.~C.~Tropper, R.~Eccleston, and C.~C.~Philips,
{\it Phys.~Rev.~B} {\bf 44}, 11 345 (1991);
M.~Seck, M.~Potemski, and P.~Wyder,
{\it Phys.~Rev.~B} {\bf 56}, 7422 (1997).

\item{[13]}
X.~M.~Chen and J.~J.~Quinn,
{\it Phys.~Rev.~Lett.} {\bf 70}, 2130 (1993);
{\it Phys.~Rev.~B} {\bf 50}, 2354 (1994);
A.~W\'ojs and J.~J.~Quinn,
{\it Phys.~Rev.~B} {\bf 63}, 045304 (2001);
A.~W\'ojs and J.~J.~Quinn,
{\it Solid State Commun.} {\bf 118}, 225 (2001).

\item{[14]}
M.~Hayne, C.~L.~Jones, R.~Bogaerts, C.~Riva, A.~Usher, 
F.~M.~Peeters, F.~Herlach, V.~V.~Moshchalkov, and M.~Henini,
{\it Phys.~Rev.~B} {\bf59}, 2927 (1999).

\item{[15]}
A.~W\'ojs, I.~Szlufarska, K.~S.~Yi, and J.~J.~Quinn,
{\it Phys.~Rev.~B} {\bf 60}, R11273 (1999).

\item{[16]}
J.~J.~Quinn, A.~W\'ojs, K.~S.~Yi, and I.~Szlufarska, 
{\it Physica E} {\bf 11}, 209 (2001).

\item{[17]}
V.~M.~Apalkov and E.~I.~Rashba,
{\it Phys.~Rev.~B} {\bf 46}, 1628 (1992);
{\it Phys.~Rev.~B} {\bf 48}, 18312 (1993).

\item{[18]}
C.~Kallin and B.~I.~Halperin,
{\it Phys.~Rev.~B} {\bf 30}, 5655 (1984).

\item{[19]}
N.~R.~Cooper,
{\it Phys.~Rev.~B} {\bf 55}, R1934 (1997).

\item{[20]}
V.~Melik-Alaverdian, N.~E.~Bonesteel, and G.~Ortiz,
{\it Phys.~Rev.~B} {\bf 60}, R8501 (1999).

\item{[21]}
S.~Das Sarma and P.~I.~Tamborenea,
{\it Phys.~Rev.~Lett.} {\bf 73}, 1971 (1994);
L.~Brey,
{\it Phys.~Rev.~Lett.} {\bf 81}, 4692 (1998);
D.~C.~Marinescu, J.~J.~Quinn, and G.~F.~Giuliani,
{\it Phys.~Rev.~B} {\bf61}, 7245 (2000).

\item{[22]}
A.~W\'ojs, P.~Hawrylak, and J.~J.~Quinn,
{\it Phys.~Rev.~B} {\bf60}, 11661 (1999).

\item{[23]}
L.~Brey, H.~A.~Fertig, R.~C\^ot\'e, and A.~H.~MacDonald,
{\it Phys.~Rev.~Lett.} {\bf75}, 2562 (1995);
R.~C\^ot\'e, A.~H.~MacDonald, L.~Brey, H.~A.~Fertig,
S.~M.~Girvin, and H.~T.~C.~Stoof;
{\it Phys.~Rev.~Lett.} {\bf78}, 4825 (1997).
B.~Paredes and J.~J.~Palacios,
{\it Phys.~Rev.~B} {\bf60}, 15570 (1999).
\end